\documentclass[twoside]{article}
\usepackage{fleqn,espcrc2}
\usepackage{amsmath}
\usepackage{graphicx} 

\title{Numerical Quantum Field Theory on the Continuum and
a New Look at Perturbation Theory%
\thanks{Presented by G.  Guralnik at {\it Lattice '01}, Berlin.}}

\author{P. Emirda\u{g}\address{Department of Physics, Brown University, Providence, RI, USA 02912--1843}, 
R. Easther\address{Institute for Strings, Cosmology and Astroparticle
Physics, Columbia University, New York, NY, USA 10027},
G. S. Guralnik$^{\mathrm a}$, S. C. Hahn$^{\mathrm a}$,
D. Petrov$^{\mathrm a}$}

\begin{document}
\begin{abstract}
\noindent The Source Galerkin method finds approximate solutions to the
functional differential equations of field theories in the presence of
external sources.  While developing this process, it was recognized that
approximations of the spectral representations of the Green's functions
by Sinc function expansions are an extremely powerful calculative tool.
Specifically, this understanding makes it not only possible to apply the
Source Galerkin method to higher dimensional field theories, but also
leads to a new approach to perturbation theory calculations in scalar
and fermionic field theories. This report summarizes the methodologies
for solving quantum field theories with the Source Galerkin method and
for performing perturbation theory calculations using Sinc
approximations.
\end{abstract}
\maketitle

\section{SOURCE GALERKIN METHOD}
\noindent To analyze a theory, we start with a Lagrangian in the presence of
external sources.  The functional differential equations satisfied by
$Z$ are generated in the usual way. The Source Galerkin technique is a
systematic, iterative algorithm to solve these equations to increasing
accuracy.  The lowest order solution is given by an ans\"atz which is an
expansion over the external sources of the theory,
\begin{equation*}
Z^{\ast}= \exp\left(\int j_x G_{xy}j_y + \int j_{\omega}j_x H_{\omega
xyz}j_y j_z + \cdots \right)
\end{equation*}
We choose $G_{xy}$ and $H_{\omega xyz}$ so that the residual associated
with the equation(s) of motion, $\hat{E}Z^{\ast}=R(j)$, is as small as
possible. We do this systematically by using weighted residuals~\cite{FLE}. We
introduce a set of test functions $\left\{j(x), j(x)j(y),
j(x)j(y)j(z),\dots\right\}$ and the measure
\begin{equation*}
[dj]\,\exp\left[\int dx\, j(x)^2/\epsilon^2\right].
\end{equation*}
We determine parameters of our expansion by imposing the constraints,
\begin {equation*}
\int[dj]\,e^{\left[j^2/\epsilon^2\right]}j(x_{i}) \cdots
j(x_{j})\left[\hat{E}Z^{\ast}\right]=0,  
\end {equation*}
where the number of constraints matches the number of free parameters.
If $Z^{\ast}$ is equal to the exact generating functional, $Z$, then
these conditions are always true.  

The integrals we encounter are of the form
\begin{equation*}
\int [dj] e^{j^2/\epsilon^2} P(j)=\int [dj] P\left(\frac{\partial}{\partial B}\right)e^{\left[-\frac{j^2}{\epsilon^2}+jB\right]} \Bigg{|}_{B=0}
\end{equation*}
Fermions are dealt with in essentially the same way, using
anticommuting sources \cite{GL1}. This method can be used in two or fewer
dimensions, but memory requirements become outrageous in higher
dimensions \cite{HG1,PEGH2}.  To surmount this obstacle, the
problem needs to be broken into smaller pieces and we must utilize the
symmetries of the action.  We do this by rewriting the usual two point
Green's function, observing that any exact two point function can be
represented as a sum of free two point functions:
\begin {equation*}
G(x,y)=\int dk^2 \, G_{{\rm free}}(k^2,x-y)B(k^2).
\end {equation*}
Introducing $\Lambda$ as a cutoff, these are
represented as
\begin {equation*}
G_{{\rm free}}(x)=\frac{1}{(2 \pi)^d} \int
dp\,\frac{e^{-p^2/\Lambda^2+ipx}}{p^2+m^{*^2}}.
\end {equation*}
Using Sinc functions \cite{STE,ET1}, we can write this in a more
numerically tractable form,
\begin {align*}
G_{\rm free}(x) &\approx G_{0}\sum\limits_{k=-N}^{N} c(k) \exp
\left[-\frac{x^{2}}{4(e^{kh} +1/ \Lambda^{2})} \right], \\
c(k) &= \frac{1}{e^{kh}}\left[\frac{\pi}{e^{kh}+1/\Lambda^{2}}
\right]^{d/2}\exp\left[ -e^{kh}m^{*^2} \right]
\end {align*}
Reducing the free propagator to sums over Gaussians suggests a
parameterization of $Z^{*}$. We write $\ln Z^{*}$ as a Feynman sum of
graphs with external legs replaced by sources but, unlike a perturbation
theory expansion, the free propagators have arbitrary mass and normalization.
The method has been applied to various field theories on the lattice
and on the continuum.  The ans\"atz is Poincar\'e invariant and will obey
other symmetries of the theory by construction, giving ample freedom
to span the solution space.  The procedure can be summarized as
follows:
\begin{enumerate}
\item Start with an ans\"atz solution $Z^{\ast}$ and derive the Source
Galerkin equations by inserting the ans\"atz into the Schwinger--Dyson
equations, and set the parameters.  This involves functional and
space-time integrals and produces nonlinear equations.
\item To improve the ans\"atz, expand $ Z^{\ast}$ with more terms and
``iterate''.
\end{enumerate}
A full account of the Source Galerkin approach is currently in
preparation.  We have made significant progress beyond the initial
lattice implementation of Source Galerkin method \cite{SGL94,GL2}. We
now can calculate on the continuum. Applications to the $\Phi^4$ in $D =
1$ and $2$, and four fermion calculations \cite{HG1,HG2} give excellent
results, even with simple $Z^*$'s which include only two terms.  The
two-dimensional nonlinear $\sigma$-model \cite{PEGH2} results are
compared to $1/N$ calculations and the leading order $\beta$ function
calculations are identical. Moreover, treating this model provided a
laboratory for investigating renormalization issues in Source Galerkin
calculations.  A study of the symmetry-breaking sector of Higgs model is
planned for the future.

\section{PERTURBATION THEORY}
\noindent The Sinc function representation used above also leads to a
new method for numerical evaluation of Feynman diagrams in conventional
perturbation theory.  In turn, the perturbative analysis is a powerful
guide to ``exact'' Source Galerkin calculations.  This approach uses
generalized Sinc functions to approximate the propagators as infinite
sums whose spatial or momentum dependence appears only in terms like
$\exp(-x^2)$.  Consequently, all integrations over internal momenta or
vertex locations are reduced to Gaussian integrals, which can be
performed analytically. To compute the diagram one has to numerically
evaluate a rapidly converging sum with dimensionality equal to the
number of propagators in the graph.  The Sinc function representation
has been applied to scalar and fermionic field theories in
\cite{ET1,ET2,PetrovET2000a}. Here we review the application of this
approach to QED calculations.

As shown in \cite{PetrovET2000a}, the fermion propagator is approximated
as
\begin {small}
\begin {eqnarray*}
S_ {Fh}(p,\Lambda)=\frac {h} {M^2}\sum\limits^ {+\infty}_
{k_1=-\infty}\exp {\left( k_1h-e^ {k_1h}\frac {m^2} {M^2} \right
)}\nonumber\\
\times(m-\gamma^\mu p_\mu)\exp {\left(-\frac {p^2} {M^2}\alpha(k_1) \right)}
\label {fp_mt}
\end {eqnarray*}
\end {small}
while the Feynman gauge photon propagator is
\begin{small}
\begin{eqnarray*}
D^ {\mu\nu}_h(p,\Lambda)=\delta^ {\mu\nu}\frac {h} {M^2} \sum\limits_
{k_1=-\infty}^ {+\infty}\exp {\ \left( k_1h-e^ {k_1h}\frac {m_v^2}
{M^2} \right) }\nonumber\\ \times\exp {\left( -\frac {p^2}
{M^2}\alpha(k_1)\right)}.
\end{eqnarray*}
\end{small}%
The accuracy and speed of any computation performed using these
expressions are governed by three factors. First, some error is
introduced through the Sinc function method itself. This error
decreases exponentially with $h$ and for $h \rightarrow 0$ the
approximation becomes exact. Additional error comes from the
truncation of the infinite sums in the approximate versions of the
propagators.  The general term of the sum, and thus the truncation
error, decreases exponentialy with the summation parameter
$k$. Finally, finite numerical precision limits the maximum accuracy
achievable in any computation. Setting a value of $h$ fixes the
expected precision of the result. Since both errors due to truncation
of sums and the Sinc expansion decrease exponentially with
corresponding parameters, the accuracy of the whole computation
improves exponentially as $h$ decreases while amount of work necessary
to perform the summations with required precision increases linearly.

In order to test this method we have evaluated several second and
fourth order diagrams.  The results of computation of second order
contribution to the anomalous magnetic moment of an electron are shown
in Table \ref{vertex-tbl}. This illustrates that accuracy as high as one part in
$10^{13}$ can be achieved using the method in question on 64-bit
hardware, and this limit dominates for $h \le 0.3$.  Performing
the same calculation with 32-bit accuracy yields a maximum accuracy of
one part in $10^6$, for $h=0.6$.

Our initial tests show that the Sinc representation method is capable
of producing results whose accuracy is limited only by numerical
precision of the hardware in a fraction of the time required to
obtain matching results by more conventional methods.
\begin{table}
\caption{\label{vertex-tbl}Correction to the magnetic moment calculated
with different values of $h$.}
\vspace{0.2cm}
\begin {center}
\begin {tabular} {|c|c|c|}\hline
$h$ & Correction$\times \alpha/\pi$ &
   Error$\times \alpha/\pi$ \\ \hline
0.2 & 0.4999999999949221 & $ 5.1 \times 10^ {-12}$\\ \hline
0.4 & 0.5000000000008583 & $ -8.6 \times 10^ {-13}$\\ \hline
0.6 & 0.500000008001 & $ -8.0 \times 10^ {-9}$\\ \hline
0.8 & 0.499999900285 & $  9.97 \times 10^ {-8}$\\ \hline
1.0 & 0.499998 & $  2.3 \times 10^ {-6}$\\ \hline
1.2 & 0.49977 & $ 0.00023 $\\ \hline
1.4 & 0.4987 & $ 0.0012 $\\ \hline
\end {tabular}
\end {center}
\end {table}
\section{SUMMARY}
The Source Galerkin method promises to tackle a range of problems
which are difficult or impossible to formulate using conventional Monte Carlo
approaches. Our methods work on the continuum and avoid most of the
usual fermionic problems. They are well suited for calculation in any
phase of a theory and are not susceptible to the ``sign problem''.

Sinc function representations can be used as a universal tool for
performing conventional perturbation theory calculations.  Accuracy as
high as one part in $10^{15}$ was achieved during testing. The method
is extremely fast and can be used to quickly obtain estimates of the
final result. High precision computations done using the Sinc
representation take significantly less time than comparable
calculations performed by with Monte Carlo methods.  We are working on
the tasks of automating this method and applying it to higher order
diagrams, as well as higher iterations of Source Galerkin.

Computational work was performed at the Theoretical Physics Computing
Facility at Brown University and the National Energy Research Scientific
Computing Center.
This work is supported by DOE contract DE--FG0291ER40688, Tasks A and D.                
\bibliography{emirdag}

\end {document}